\documentclass[aps,preprint,showpacs, showkeys]{revtex4}%
\usepackage{amsfonts}
\usepackage{amsmath}
\usepackage{amssymb}
\usepackage{graphicx}%
\providecommand{\U}[1]{\protect\rule{.1in}{.1in}}

\begin{document}
\title[ ]{Reply to "Comment on 'New ansatz for metric operator calculation in
pseudo-Hermitian field theory'~"}
\author{Abouzeid M. Shalaby}
\email{amshalab@ mans.edu.eg}
\affiliation{Physics Department, Faculty of Science, Mansoura University, Egypt.}

\begin{abstract}
In this report, we reply to a recent comment by Carl M. Bender, Gregorio
Benincasa and Hugh F. Jones on our work 'New ansatz for metric operator
calculation in pseudo-Hermitian field theory (Phys. Rev. D. 79, 107702
(2009)). In fact, they figured out that there exist sign errors in our work
which leaded to the conclusion of the invalidity of the ansatz introduced in
our work. Here, we show that, the ansatz is valid in d+1 space-time
dimensions, which by itself is a new and very important result. The importance
of the work comes from the fact that it is the first time to have a metric
operator for a quantum field theory which is local in the fields as well as
valid in $3+1$ space-time dimensions. Moreover, it is composed of the
operators in the Hamiltonian itself which makes the Feynmann diagram
calculations for the physical amplitudes go the same way as in conventional theories.

\end{abstract}
\maketitle

In a recent comment, Carl M. Bender, Gregorio Benincasa and Hugh F. Jones
\cite{benderans} found out some sign errors in our work in Ref.\cite{abomet}
which leaded them to conclude that the ansatz introduced in our work in
Ref.\cite{abomet} does not work in any space-time dimensions. In this report,
we will show that the ansatz is valid not only in $1+1$ space-time dimensions
but also in all the space-time dimensions and represents the first
calculations for a metric operator in the real world for a non-trivial
non-Hermitian field theory.

The calculations of the metric operator are indispensable for any physical
calculations in a pseudo-Hermitian theory. \ An important property of the
metric operator is that it is not unique
\cite{unique00,unique0,unique1,unique2,unique3,unique4}. \ This gives us the
freedom to conjecture a simple form that can be employed in the calculations
of the physical amplitudes in a practical way. In Ref. \cite{bendmet}, a form
of the $C$ operator is obtained for the $i\phi^{3}$ field theory in $1+1$
dimensions. This form is non-local in the fields and includes complicated
parameters. In our work in Ref. \cite{abomet}, we introduced a new ansatz for
the metric operator which is local in the fields and of fixed parameters as
well. The idea of the ansatz is that the metric operator is conjectured to be
a functional of all the operators in the Hamiltonian. To advocate this nice
and interesting idea, in the following we present the final results of the
first order $Q$ operator for the $i\phi^{3}$ field theory. Note that, in the
following, when two gradient operators appears in some formula this means that
there exists an inner product between them \textit{i.e} $\nabla\nabla
\equiv\nabla\cdot\nabla$.

Now, consider the Hamiltonian model of the form;%
\begin{equation}
H=\frac{1}{2}\left(  \left(  \nabla\phi\right)  ^{2}+\pi^{2}+m^{2}\phi
^{2}\right)  +ig\phi^{3}, \label{ham}%
\end{equation}
where $\phi$ is the field operator, $\pi$ is the conjugate momentum and $g$ is
the coupling constant. The metric operator can be assumed to have the form
$\eta=\exp\left(  -Q\right)  $, where $Q$ is Hermitian. Note that, although
the Hamiltonian in Eq.(\ref{ham}) is non-Hermitian in a Hilbert space with the
Dirac sense inner product, it is Hermitian in a Hilbert space endowed by the
inner product $\langle n|\eta|n\rangle$.

To obtain $Q$, we rename the terms in the Hamiltonian as;
\[
H=H_{0}+gH_{I}%
\]%
\begin{align*}
H_{0}  &  =\frac{1}{2}\int d^{d}x\left(  \left(  \nabla\phi\right)  ^{2}%
+\pi^{2}+m^{2}\phi^{2}\right)  ,\\
H_{I}  &  =i\int d^{d}x\phi^{3},
\end{align*}
$d$ is the spatial dimension. In using the relation $H^{\dagger}=\eta
H\eta^{-1}$, we get%

\begin{align*}
H^{\dagger}  &  =\exp(-Q)H\exp(Q)=H+[-Q,H]+\frac{[-Q,[-Q,H]]}{2!}\\
&  +\frac{[-Q,[-Q,[-Q,H]]]}{3!}+....,
\end{align*}
where%
\[
Q=Q_{0}+gQ_{1}+g^{2}Q_{2}++g^{3}Q_{3}+..
\]
Up to first order in the coupling, we have the operator equation;%

\begin{equation}
-2H_{I}=\left[  -Q_{1},H_{0}\right]  . \label{frstord}%
\end{equation}
In Ref.\cite{abomet}, we conjectured \ $Q_{1}$ to have the form;%

\begin{align}
Q_{1}  &  =C_{1}\int d^{d}z\pi^{3}(z)+\frac{C_{2}}{3}\int d^{d}z\left(
\pi(z)\phi^{2}(z)+\phi(z)\pi(z)\phi(z)+\phi^{2}(z)\pi(z)\right) \nonumber\\
&  +\frac{C_{3}}{3}\int d^{d}z\left(  \pi(z)\nabla\phi(z)\nabla\phi
(z)+\nabla\phi(z)\pi(z)\nabla\phi(z)+\nabla\phi(z)\nabla\phi(z)\pi(z)\right)
.
\end{align}
To get the parameters $C_{1}$, $C_{2}$ and $C_{3}$, we substitute this form of
$Q_{1}$ in Eq. (\ref{frstord}).

Now, in commuting the first term in $Q_{1}$ with the second term in $H_{0}$ we get;%

\begin{align}
&  C_{1}\int d^{d}x\int d^{d}z\left[  \pi^{3}(z),\frac{1}{2}\nabla_{x}%
\phi\left(  x\right)  \nabla_{x}\phi\left(  x\right)  \right] \nonumber\\
&  =\frac{3i}{2}C_{1}\int d^{d}x\pi^{2}(x)\nabla_{x}^{2}\phi\left(  x\right)
+\frac{3i}{2}C_{1}\int d^{d}x\nabla_{x}^{2}\phi\left(  x\right)  \pi
^{2}(x),\nonumber
\end{align}
where it differers from the result in Ref.\cite{abomet} by an overall negative sign.

For the first term in $Q_{1}$with the third in $H_{0}$, we have%

\begin{align}
&  \frac{1}{2}m^{2}C_{1}\int d^{d}x\int d^{d}z\left[  \pi^{3}(z),\phi\left(
x\right)  \phi\left(  x\right)  \right] \nonumber\\
&  =\frac{-3i}{2}m^{2}C_{1}\int d^{d}x\pi^{2}(x)\phi\left(  x\right)
+\frac{-3i}{2}m^{2}C_{1}\int d^{d}x\phi\left(  x\right)  \pi^{2}(x).\nonumber
\end{align}

Now, for the second term in $Q_{1}$with the first in $H_{0}$, we then have;
\begin{align}
&  \frac{C_{2}}{6}\int d^{d}x\int d^{d}z\left[  \pi(z)\phi^{2}(z)+\phi
(z)\pi(z)\phi(z)+\phi^{2}(z)\pi(z),\pi^{2}(x)\right] \nonumber\\
&  =iC_{2}\int d^{d}x\phi(x)\pi^{2}(x)+iC_{2}\int d^{d}x\pi^{2}(x)\phi
(x)\nonumber
\end{align}

Also, for the second term in $Q_{1}$with the second in $H_{0}$, we have;
\begin{align}
&  \frac{C_{2}}{6}\int d^{d}x\int d^{d}z\left[  \pi(z)\phi^{2}(z)+\phi
(z)\pi(z)\phi(z)+\phi^{2}(z)\pi(z),\nabla_{x}\phi\left(  x\right)  \nabla
_{x}\phi\left(  x\right)  \right] \nonumber\\
&  =iC_{2}\int d^{d}x\phi^{2}(x)\nabla_{x}^{2}\phi\left(  x\right)  ,
\label{T22}%
\end{align}
where it differers from the result in Ref.\cite{abomet} by an overall negative sign.

For the second term in $Q_{1}$with the third in $H_{0}$, we have%

\begin{align}
&  \frac{C_{2}m^{2}}{6}\int d^{d}x\int d^{d}z\left[  \pi(z)\phi^{2}%
(z)+\phi(z)\pi(z)\phi(z)+\phi^{2}(z)\pi(z),\phi\left(  x\right)  \phi\left(
x\right)  \right]  ,\nonumber\\
&  =-iC_{2}m^{2}\int d^{d}x\phi^{3}\left(  x\right)  . \label{T23}%
\end{align}

For the commutator of the third term in $Q_{1}$ with the first in $H_{0}$;

\bigskip\ %

\begin{align}
&  \frac{C_{3}}{6}\int d^{d}x\int d^{d}z\left[  \nabla_{z}\phi(z)\nabla
_{z}\phi(z)\pi(z)+\nabla_{z}\phi(z)\pi(z)\nabla_{z}\phi(z)+\pi(z)\nabla
_{z}\phi(z)\nabla_{z}\phi(z),\pi^{2}(x)\right]  ,\nonumber\\
&  =-iC_{3}\int d^{d}x\pi(x)\pi(x)\nabla^{2}\phi(x)-iC_{3}\int d^{d}%
x\nabla^{2}\phi(x)\pi(x)\pi(x). \label{T31}%
\end{align}

For the third of $Q_{1}$ with the second in $H_{0}$ we get;
\begin{align}
&  \frac{C_{3}}{6}\int d^{d}x\int d^{d}z\left[
\begin{array}
[c]{c}%
\nabla_{z}\phi(z)\nabla_{z}\phi(z)\pi(z)+\nabla_{z}\phi(z)\pi(z)\nabla_{z}%
\phi(z)\\
+\pi(z)\nabla_{z}\phi(z)\nabla_{z}\phi(z),\nabla_{x}\phi(x)\nabla_{x}\phi(x)
\end{array}
\right] \nonumber\\
&  =iC_{3}\int d^{d}x\nabla^{2}\phi(x)\nabla\phi(x)\nabla\phi(x). \label{T32}%
\end{align}

Also, for the third of $Q_{1}$ with the third in $H_{0}$ we get;

then%
\begin{align}
&  \frac{m^{2}C_{3}}{6}\int d^{d}x\int d^{d}z\left[
\begin{array}
[c]{c}%
\nabla_{z}\phi(z)\nabla_{z}\phi(z)\pi(z)+\nabla_{z}\phi(z)\pi(z)\nabla_{z}%
\phi(z)\\
+\pi(z)\nabla_{z}\phi(z)\nabla_{z}\phi(z),\phi(x)\phi(x)
\end{array}
\right] \nonumber\\
&  =\frac{im^{2}C_{3}}{2}\int d^{d}x\nabla^{2}\phi(x)\phi^{2}(x). \label{T33}%
\end{align}

For Eq.(\ref{frstord}) to be satisfied, we have the following algebraic
equations;%
\begin{align*}
\frac{3i}{2}C_{1}-iC_{3}  &  =0,\\
\frac{-3i}{2}m^{2}C_{1}+iC_{2}  &  =0,\\
-iC_{2}m^{2}  &  =2i,\\
iC_{2}+\frac{im^{2}C_{3}}{2}  &  =0.
\end{align*}
Moreover, we have the integral $\int d^{d}x\nabla^{2}\phi(x)\nabla
\phi(x)\nabla\phi(x)$ which has to vanish as well. In fact, the above set of
equations are inconsistent as they are four different equations in three
unknowns. Nevertheless, one can verify the validity of the ansatz. To show
this, we consider the fact that the quantized fields have to satisfy the
klein-Gordon equation because we have the relation;
\begin{equation}
\overset{\cdot}{\pi}=-i\left[  \pi,H\right]  =\left(  \nabla^{2}-m^{2}%
-3ig\phi\right)  \phi,
\end{equation}
which is nothing but the klein-Gordon equation for the field $\phi$. This
non-linear equation can be solved exactly in the static frame with the
solution given by \footnote{The details of how to obtain this exact solution
will be presented in another work (in progress) but at the moment one can
easily check that it really verify the Klein-Gordon equation.};%

\[
\phi\left(  x,y,z\right)  =\frac{im^{2}}{2g}sech^{2}\left(  \frac{m\left(
x+y+z\right)  }{2\sqrt{3}}\right)  .
\]
Since the field $\phi\left(  x\right)  $ transforms as $\phi\left(  \Lambda
x\right)  $ under Lorentz group, one can get the exact solution at any frame
of reference. With no loss of generality, let us assume the motion to be in
the $x$-direction with a velocity $v$, then one have the solution;%
\[
\phi\left(  x,y,z,t\right)  =\frac{im^{2}}{2g}sech^{2}\left(  \frac{m\left(
\gamma\left(  x-vt\right)  +y+z\right)  }{2\sqrt{3}}\right)  ,
\]
where $\gamma$ is the Lorentz factor. This solution has the following
properties;%
\begin{align*}
\nabla^{2}\phi &  =\left(  3-2v^{2}\right)  \frac{\partial^{2}\phi}{\partial
x^{2}},\\
\frac{\partial\phi}{\partial y} &  =\frac{\partial\phi}{\partial z}%
=\sqrt{1-v^{2}}\frac{\partial\phi}{\partial x},\\
\pi &  =\frac{\partial\phi}{\partial t}=-v\frac{\partial\phi}{\partial x}.
\end{align*}
Accordingly we have; \
\begin{align*}
\nabla^{2}\phi(x)\left(  \nabla\phi(x)\right)  ^{2}  & =\left(  v^{2}%
-2\right)  \left(  -3+2v^{2}\right)  \left(  \frac{\partial\psi}{\partial
x}\right)  ^{2}\left(  \allowbreak\frac{\partial^{2}\psi}{\partial x^{2}%
}\right)  ,\\
\nabla^{2}\phi(x)\pi^{2}(x)  & =\left(  3-2v^{2}\right)  v^{2}\left(
\frac{\partial\psi}{\partial x}\right)  ^{2}\left(  \allowbreak\frac
{\partial^{2}\psi}{\partial x^{2}}\right)  ,
\end{align*}
and since the integral $\int dx\frac{\partial^{2}\phi(x)}{\partial x^{2}%
}\left(  \frac{\partial\phi(x)}{\partial x}\right)  ^{2}$ is a vanishing
surface term then the integrals $\int d^{d}x\nabla^{2}\phi(x)\nabla
\phi(x)\nabla\phi(x)$ and $\int d^{d}x\nabla^{2}\phi(x)\pi^{2}(x)$ also
vanish. Accordingly, for Eq.(\ref{frstord}) to be satisfied we only have the
following three algebraic equations for the parameters $C_{1}$, $C_{2}$ and
$C_{3}$;%
\begin{align*}
\frac{-3i}{2}m^{2}C_{1}+iC_{2} &  =0,\\
-iC_{2}m^{2} &  =2i,\\
iC_{2}+\frac{im^{2}C_{3}}{2} &  =0.
\end{align*}
In other words, we have the parameters values as $C_{1}=-\frac{4}{3m^{4}%
},C_{2}=-\frac{2}{m^{2}},C_{3}=\frac{4}{m^{4}}$, which is the same result
reported in our work in Ref. \cite{abomet} except that $C_{3}$ has reversed
its sign.

\bigskip

\bigskip

To conclude, we have shown that the ansatz introduced in our work
\cite{abomet} is applicable in any space-time dimensions although of the sign
errors mentioned in Ref.\cite{benderans}. To show this, we relied on the
properties of the exact form of the field which assured the validity of the
ansatz. In fact, our ansatz is an important starting point for a concrete
formulation of pseudo-Hermitian field theories since it is local in the fields
and of simple shape. \ Moreover, the ansatz represents the first calculation
for the metric operator of a non-trivial pseudo-Hermitian scalar field theory
in the real world of $3+1$ space-time dimensions.

\end{document}